\documentclass{llncs}
\usepackage[T1]{fontenc}
\usepackage{graphicx}
\begin{document}
\title{The Application of Procedurally Generated Libraries in Immersive Virtual Reality}
%
%
\title{The Application of Procedurally Generated Libraries in Immersive Virtual Reality}
%
%
\author{Saeed Safikhani \inst{1}\and
Benedikt Gross \inst{1}\and
Johanna Pirker \inst{1,2}}

\institute{Graz University of Technology, Graz, Austria \and 
Ludwig Maximilian University, Munich, Germany}

\authorrunning{S. Safikhani et al.}
%
%
\maketitle              
\begin{abstract}
While digital libraries offer essential benefits for the digital age such as constant availability and accessibility, they often fail in providing a similarly enjoyable browsing experience that benefits engagement and serendipitous exploration inherent to traditional library experiences. In this paper, we propose a virtual reality (VR) library environment that emulates a traditional library promoting an engaging browsing experience that benefits browsing enjoyment and serendipity. The implemented system uses procedural content generation to dynamically generate a library-like environment, wherein each room corresponds to a specific category of the digital library. To assess the usability of the implemented system, we conducted an A/B study comparing the VR environment with the web interface of \textit{Project Gutenberg}. The results of the study suggest that while the physical demand for using the VR environment is higher, the system can significantly benefit the browsing experience and entice users’ curiosity. Furthermore, the study suggests that the system can increase user engagement and users generally felt that using the system was rewarding.

\end{abstract}
\section{Introduction}

Since the inception of digital libraries, significant contributions from diverse fields, including library and computer science, have advanced this research area \cite{Chowdhury2003}. Notably, extensive digitalization initiatives like the \textit{Google Book Project} have substantially increased the availability of digital resources \cite{XieIris2016}. 
In contemporary times, digital libraries have gained paramount importance, playing a crucial role in disseminating knowledge and information, as underscored by the COVID-19 pandemic \cite{pambayun2021}. Despite these advancements, conventional interfaces of digital libraries often fall short in facilitating effective browsing and serendipitous discoveries compared to their physical counterparts \cite{LIU2011}. Traditional libraries leverage visual recognition and spatial reasoning, allowing users to discover thematically related items within the same room or on adjacent bookshelves \cite{Kleiner2013}. With the increasing prevalence and affordability of virtual reality (VR) headsets, VR technology offers an immersive and interactive solution for enhancing user experiences \cite{Wohlgenannt2020}. Leveraging VR for digital libraries presents a unique opportunity to emulate the traditional library environment, utilizing the entire body for book discovery in a manner that aligns with familiar physical library experiences \cite{Cook2018}. This immersive approach can potentially address the limitations of current digital library interfaces and enhance the browsing experience and serendipity.

The main contribution of this paper is to create an immersive VR library system that mirrors the experience of navigating a physical library, providing users access to a diverse range of books. The system's environment will be procedurally generated for easy adaptation to different book catalogs, ensuring flexibility and scalability. Users will have the opportunity to read through the entire library collection within an immersive setting, supplemented by additional information and summaries generated using a language completion model. The system will also incorporate exhibition objects related to different library categories to enhance exploration and stimulate curiosity. To evaluate the benefits of the system, a study was conducted to compare the VR library environment with traditional digital library interfaces, focusing on the user's browsing and reading experience.


\section{Background and Related Works}

A digital library is an organization that provides access to digital resources, including databases of text, numbers, graphics, sound, video, and other types of media. It is a service, an architecture, a set of information resources, and a set of tools and capabilities to locate, retrieve, and utilize the information resources available \cite{Walters1998}.
Due to the importance of digital libraries and repositories as infrastructure, There has been considerable research and development for them in recent years \cite{Ahmad2018}. Compared to their physical counterparts, they can provide easier access for a wide range of users. This can be especially useful when users need to access them remotely, either due to the distance to the physical resources or restriction of access to them, such as the condition during the COVID-19 pandemic \cite{pambayun2021}.
With the substantial increase of objects in digital libraries and the enormous expansion of data online, information retrieval (IR) has seen high interest in commercial applications and research communities. While not exclusive to libraries these methods offer ways of providing users with the right information and results even from messy search queries \cite{Ibrihich2022}.
The process of IR for text-based data usually consists of some pre-processing on the documents. This includes methods like tokenizing, stemming, stop word removal, and some form of feature extraction. Based on these features some retrieval models like the Boolean Model or the Vector Space Model are applied to find appropriate search results \cite{manning2009}.
IR not only but especially in libraries also heavily relies on cataloging and metadata to provide additional information when searching for books. This particularly applies to Digital Libraries with large collections of documents \cite{Onwuchekwa2011}.
Although providing a list of search results is an adequate solution for seeking certain information, the simple listing of relevant documents or books as done by many digital libraries can stand in the way of serendipitous discoveries often gained by traditional library shelf browsing \cite{Ford2009}.

The barrier towards serendipitous discoveries can be due to the lack of a visual representation of available books in digital libraries, which is a benefit for physical libraries \cite{LIU2011}. Browsing habits can best be examined in traditional libraries as there has only been little effort put into replicating physical shelf browsing in digital libraries \cite{unlu2021}.
Here, studies show that the lending of neighboring books increases the chances of a subsequent loan by at least 9-fold \cite{McKay2014} and users seem to be generally successful when browsing the physical stacks \cite{Mckay2017}. Furthermore, library patrons often seem to prefer the physical browsing of books over searching in digital libraries \cite{Mckay2011}. 
Although browsing is no replacement for classical IR, it can be a significant complement as it makes use of visual recognition and spatial reasoning over linguistic specification and logical reasoning \cite{poulter2003}.
\cite{hinze2012} and \cite{Mckay2017} both examined traditional library patron's book selection behaviors and present features that can be implemented in Digital Libraries to better facilitate these habits. Some of these features are \textit{seamless transitions, support of zooming while browsing, comparison of books, and Displaying a large collection at once.}
Many of these suggestions for digital library visualizations have already been applied in implementations by either directly emulating the physical library experience digitally, e.g.~\cite{Lemons2019}, or single bookshelf \cite{unlu2021}.
The first implementation is mainly called \textit{Digital Library Replicas}, where \textit{Blended Shelf} is used for the latter.
The Blended Shelf is a digital interface developed in response to the archiving of books from the University of Konstanz's library due to an asbestos discovery. It mimics physical bookshelves, allowing users to select disciplines and browse corresponding books with accurate representations. Presented on an interactive whiteboard, users can view book covers without direct interaction. In contrast, the Library of Apollo is a web-based e-reader offering access to over 9.4 million digitized books from 200,000 categories, enabling browsing and search functionality \cite{Kleiner2013,unlu2021}.
Emulating the physical library may lead to preserving browsing and serendipitous discovery in digital format as it can provide an already familiar browsing experience.

The overgrowing application of immersive technologies shows new opportunities for visualization and interaction in virtual environments \cite{Slater2016}. These immersive experiences can be a solution for visualization in different scenarios such as digital libraries.
Cook et al. \cite{Cook2018} suggests that this technology especially lends itself to preserving browsing and serendipitous discovery in digital format as it can integrate the whole body, therefore, providing an already familiar browsing experience.
In this case, VR can be a solution to improve the representation of digital libraries to provide serendipitous browsing and reading of books.
According to previous studies, VR has been shown to provide customizable multimedia content, higher immersion and presence, and effective information transfer in the context of cultural heritage \cite{chong2022comprehensive,berti2021unexplored}. VR technology can be particularly useful for reconstruction/digitalization and improvement in the presentation and preservation of cultural heritage sites and artifacts. This approach could also be applied to VR digital libraries for visualization in similar environments.

Although VR shows potential as a medium for digital libraries, the quality of the reading experience can play a crucial role in the user acceptance of the technology. 
When comparing reading from paper and screens the most commonly used metrics of measuring performance are investigating text comprehension, reading speed, and meta comprehension which describes how well the reader is able to make judgments about their text comprehension \cite{halamish2020}. Even though the body of studies investigating the differences in reading is large, research is often contradictory \cite{Sage2020} and while \cite{Yehudah2021} for example argue that studies have not yet come to a conclusive result, many meta-studies suggest a general inferiority of reading from screens \cite{delgado2018,kong2018,Furenes2021,clinton2019}.
Most studies compared reading plain texts digitally and on paper not making use of multimedia and other functions of digital devices that as \cite{Schwabe2022} suggest can positively affect reading comprehension. According to \cite{kaplan2022a}, virtual reality (VR) can provide an immersive experience to its users. 

Although little research has currently been done on comparing VR and paper reading, a study by \cite{Baceviciute2021} found that reading in VR can improve knowledge transfer and participants considered reading less demanding. The study also found that users often had difficulties reading small text in VR, however other studies already suggested guidelines for dealing with this issue.
An overview of the literature regarding a guideline for better reading experience in VR can be summarized as the following:
\begin{itemize}
    \item \textbf{Reading Surface:} reading on a concave surface that is only warped around one axis is preferable. This curvature can vary between 50\textdegree to 70\textdegree for surface warp \cite{Wei2020}.
    \item \textbf{Field of View:} An FoV of 90\textdegree from the HMD combined with a comfortable head movement range of 30\textdegree horizontally, 20\textdegree vertically upwards and 12\textdegree downwards provide an area for so-called the \textit{content zone} \cite{Alger2015}.
    \item \textbf{Text Height:} a general text height of 32 dmm\footnote{Distance Independent Millimetre: 1 dmm equals to a height of 1 mm when viewed at a distance of 1 m} best suited with a minimum of 23 dmm \cite{dingleer2018,Google2017}.
    \item \textbf{Text Font:} text font is mostly suggested to be kept as sans serif to provide visual clarity \cite{dingleer2018,Cavalcanti2022}.
    \item \textbf{Text Placement:} it is suggested that the reading interface is at least partially fixed as too much movement can reduce usability \cite{Rzayev21} and face slightly down \cite{Google2017}.
\end{itemize}

 Since the design and creation of a 3D model of a library can be time-consuming especially since new resources are continually added to their catalog, Procedural Content Generation (PCG) can serve as a great solution for an approach that is able to adapt automatically to new resources \cite{cogo2023survey}. PCG describes the automated creation of media such as poetry, paintings, architecture and content for games \cite{Barriga2019}. PCG can provide large amounts of content and provide new and different experiences each time while often also being able to adapt to a multitude of situations \cite{Risi2019,Short2017}.


 In this study, we used PCG to generate the content of a digital library for a VR environment. In this way, we benefit from both the immersive experience of VR and the automation functionality of PCG to maintain the library contents.

\section{Procedurally Generated VR Library}
\begin{figure}
\centering
\includegraphics[width=10cm]{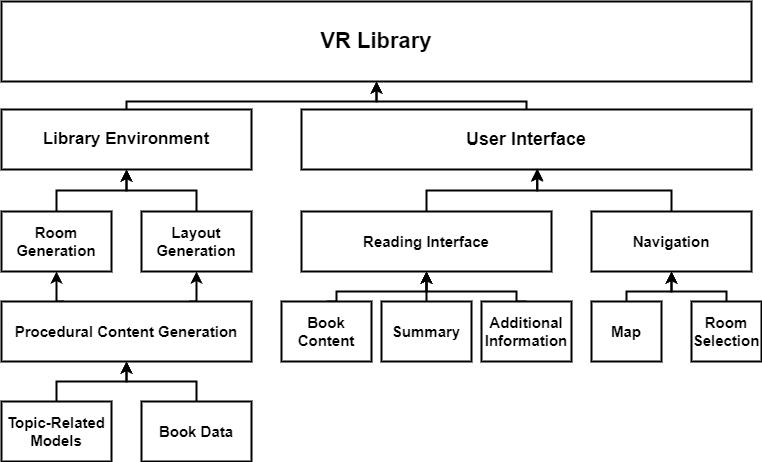}
\caption{Conceptual Architecture of the proposed VR library application.}
\label{conceptual_arch}
\end{figure}

In this study, we proposed a system to generate a digital library with customizable rooms, based on the book content of the provided library.
The conceptual architecture of this application is demonstrated in Fig.~\ref{conceptual_arch}.

Accordingly, The PCG system in this study was developed for two main purposes: (1) generation of individual rooms in the library containing the books based on their topic (2) generation of the floor layout and concatenation of rooms. 

\subsection{Procedural Room Generation}
Based on the number of books in each catalog, the PCG system for rooms should be able to set the size of the room accordingly. In addition, the room design should provide enough space for the player to be able to move freely inside the room and between bookshelves. In this way, we designed rooms as a cuboid where all rooms share the same height. We defined for PCG tool to use the edge of the room for the bookshelves. We added an entrance and exit for each to the different side of the room to make it easier for the user to be oriented in a VR environment. These design decisions define the main structure of the room. 


In addition, we added some decorative objects to the room, such as furniture and category-related 3D models, to improve the visual presentation of the environment. The 3D models are interactable in such a way that users can rotate them and read related information about them.

\subsection{Procedural Layout Generation}
Rooms have at least one entrance and exit to ensure a continuous path from the first room to the last. To ensure that rooms are in close proximity to each other, rooms are laid out in a clockwise or counter-clockwise pattern. However, since the size of each room varies, rooms might still not directly align with others. To further compress the layout, new rooms are, if possible, placed directly at the room south of the current placement direction while still being connected to the outgoing room.
When rooms are directly adjacent to each other, they can be connected, opening up a new path to traverse through the environment. The size of a room thereby determines how many connections to other rooms are possible, since too many junctions in a small space can lead to an overly complex layout and make navigation confusing.


\subsection{Navigation in the Procedurally Generated Library}

Navigation in a procedurally generated structure can be difficult and without additional information, users might become disorientated and lost.
We used signboards and a map to visualize the current player location in the virtual library layout. This map can also be used as a room selection interface, and users can teleport between rooms by selecting them either based on their location or category.
This map supports gesture-controlled scrolling and is scalable to change the UI size. The movement of the player is provided by a \textit{Continuous Move Provider} that allows for smooth movement through inputs of the left controller's d-pad and a \textit{Continuous Turn Provider} that similarly allows for rotation through the left and right inputs from the right controller's d-pad.

\subsection{Additional Reading Context}
To further improve the browsing experience AI can provide the user with additional resources about the currently selected book. These resources can include additional information and summaries. \textit{OpenAI}
provides various reinforcement learning models that are trained on a large corpus of web pages and books. Their language models provide text answers given an input prompt.
We used this language model to provide additional information related to the selected book.
The book information received from the API requests can be displayed in a similar manner as the book content. However, since the answers provided by the models from \textit{OpenAI} contain a lot less text than the book content, the page-turning can be replaced by scrollable text. In this way, simple buttons can be used to switch context between book content, additional information and book summary.

\subsection{Implementation}
\textbf{Development Environment:} We developed our system in Unity 2020.3.4f
and we used the \textit{XR Interaction Toolkit} as an interaction system for the VR implementation. The \textit{XR Interaction Toolkit} provides some valuable components such as a room-scale VR-camera-rig, capabilities for interaction with UI elements, grabbable objects, and more. This interaction system supports various controller input systems and is in this case used in combination with \textit{OpenXR}
\textit{OpenXR} provides a general application interface that allows the use of a single API for a multitude of different VR devices and systems such as SteamVR
, and Oculus
.

\textbf{Database:} We used the project Gutenberg\footnote{https://www.gutenberg.org/} as a database for our ebooks. This digital library contains over 70'000 ebooks that focuses on older works for which US copyright has expired. While other digital libraries such as the \textit{Internet Archive}\footnote{https://archive.org/details/texts} feature a more extensive catalog with over 37 million books, \textit{Project Gutenberg} shows better support for the acquisition of their provided resources and is therefore preferable for showcasing the proposed system.

\begin{figure} [!h]
    \centering
    \includegraphics[width=\textwidth]{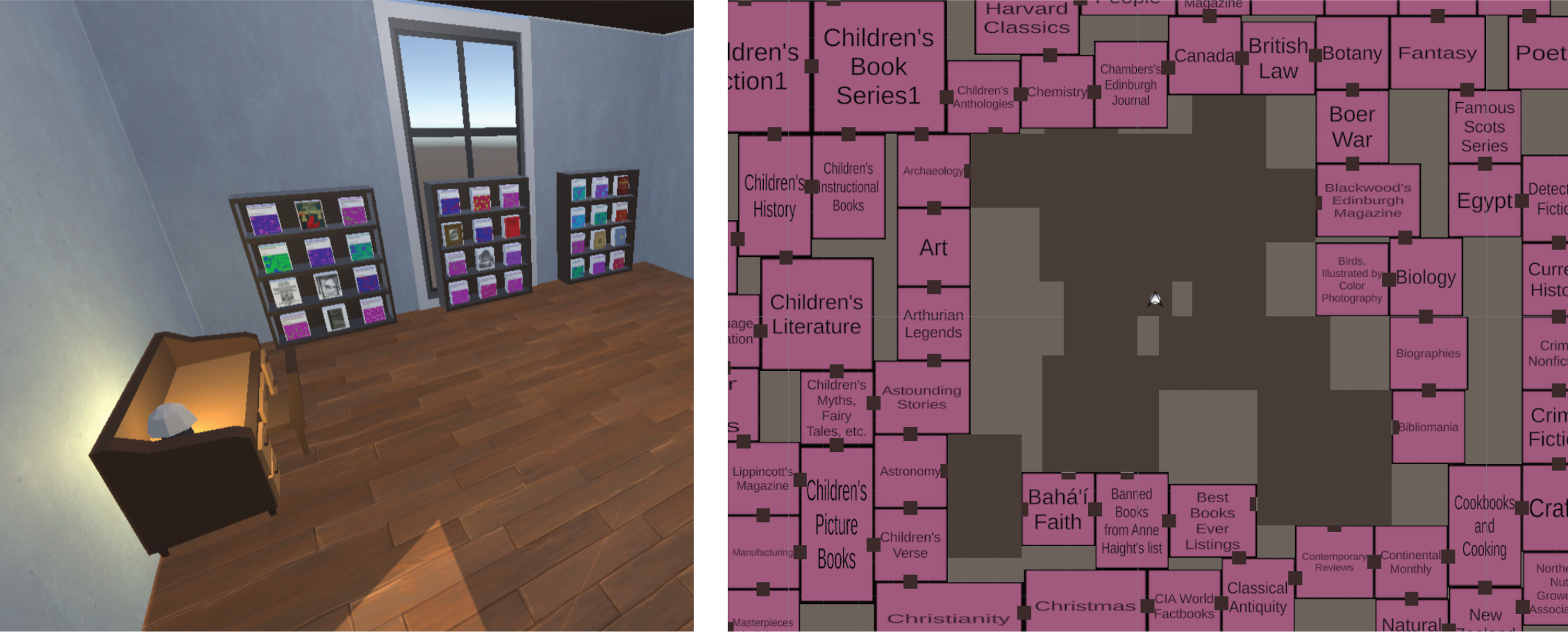}
    \caption{(left)An example of a finished room including bookshelves, windows and furniture, (right)The room concatenation in the procedurally generated library}
    \label{fig:room_layout_mix}
\end{figure}

\textbf{Library Generation:} 
Rooms are created based on how many books are present in each category. Most rooms are created with either the width or the height as a fixed value that primarily considers enough space for the player to traverse the room. The other length of the room is determined by how many bookshelves need to be placed. A newly created room is then concatenated to the previous room, see Fig.~\ref{fig:room_layout_mix} (right). After every room has its place and all connections are made, all room objects can be instantiated beginning with the room structure such as floor, walls, and ceiling. Furthermore, essential other objects such as lights, door signs, and room-dependent user interface elements are also created in this step. An example of the generated room is demonstrated in Fig.~\ref{fig:room_layout_mix} (left).

\begin{figure}[h]
    \centering
    \includegraphics[width=0.9\textwidth]{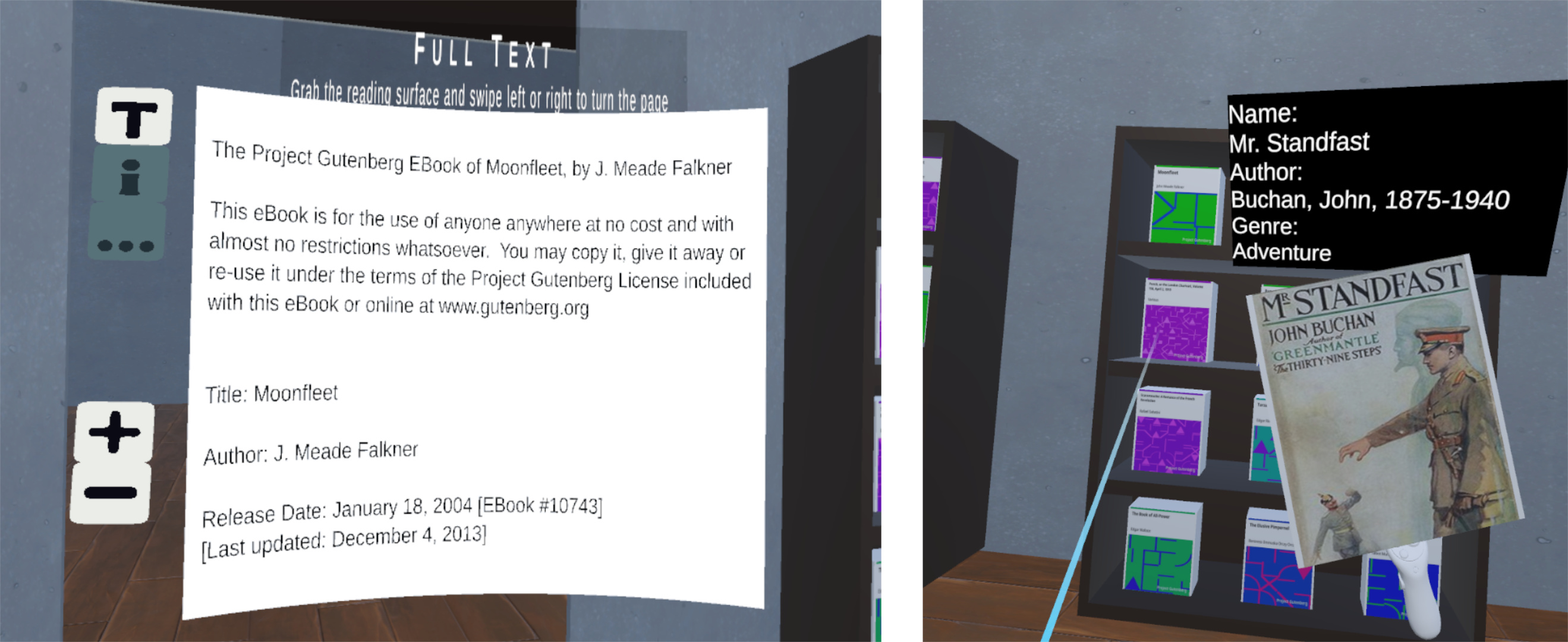}
    \caption{(left) reading interface with corresponding UI elements (right) hover and grab interaction}
    \label{fig:interactions}
\end{figure}

\textbf{Interactions:} 
Each book contains a \textit{Grab Interactable Component} that enables the \textit{Hover} and \textit{Grab} interactions. The hover interaction conveys information about the hovered book, including title, author, publication year, and category. Pressing the grab button on the hovered book picks up the book and by pressing the trigger button user can read it. These interactions are visualized in Fig.~\ref{fig:interactions}.


\subsection{Optimization}
The implemented procedural library deals with lots of data in the form of text, cover images and objects that are active at the same time. Optimization of the provided system is therefore vital to ensure adequate performance. 
While too much data can be limiting in terms of memory, rendering too many objects at once can have a large impact on performance. To counteract this problem, not all rooms and their objects are rendered at the same time. Furthermore, interior objects are only rendered when the user is directly inside a room. 

 \section{Methodology}

 In order to compare our system for procedurally generated library in VR with traditional digital library interfaces, we conducted a user study.
 The main objective of this user study is to figure out whether the immersion and interactivity of the VR library can lead to user engagement, browsing interest, and enjoyment.

\textbf{Materials and Study Design:}
The experiment was conducted as an A-B study. Group A started using the VR environment and later switched to the web interface while Group B began with the web application and then switched to the VR system. We considered a pre-questionnaire, a demographic questionnaire, to gather general participant information (e.g. age, gender, level of education) as well as their background related to the study (e.g. previous experience with visual interface, and prior experience in VR). We evaluated the user experience after each interface using post-questionnaire. This questionnaire was a collection of standard questionnaires and a custom questionnaire regarding browsing and reading experience in each interface. In this case, we considered the System Usability Scale (SUS) \cite{Brooke1995}, the User Engagement Scale (UES)  \cite{Obrien2018}, and the NASA Task Load Index (NASA TLX) \cite{Hart1988} as standard questionnaires.

\textbf{Setup:}
We used a PC with an Intel Core i5 CPU, Nvidia RTX 3080Ti GPU, and 32 GB RAM to provide a smooth VR experience. For VR setup, we considered 
"Meta Quest 2"\footnote{https://www.meta.com/at/quest/products/quest-2/}. In this way, we were able to run the study without predefined tracking sensors in the room. 

\textbf{Participants:}
During the course of the study, 20 people participated with 15 being male and 5 female. The participants were divided equally into two groups each starting with a different system. Group A started with the assessment of the VR environment and group B evaluated the Project Gutenberg web system first. The ages of participants ranged from 16 to 63 (AVG=26.65, STD: 9.43). Out of the 20 participants, 9 completed a bachelor’s degree, 2 a master’s degree, and 9 with lower education level. Participants answered to regularly use a computer (AVG: 4.70/5.0, STD: 0.73) and rated themselves as high for adjusting to new interfaces (AVG: 4.05/5.0, STD: 0.89).  Rating of experience with video games was mixed (AVG: 3.40/5.0, STD:1.64) and participants rated themselves low when it comes to previous experience with virtual reality (AVG: 1.80/5.0, STD: 0.89).

\textbf{Procedure:}
The study for each participant began by filling out pre-questionnaires. After this step, according to the assigned group (A or B), they continued with the corresponding system (procedural in VR or traditional digital). For VR groups, as it could be the first time for them to experience a VR environment, we explained to them the general controller schema as well as UI functionality. They were asked to fulfill the following tasks to complete each scenario:

\begin{enumerate}
    \item Find a book in the category Harvard Classics that you find interesting.
    \item Find a book you would be interested in reading in a category about a country.
    \item Find a book you would be interested in reading in a category of your choice.
\end{enumerate}

Since the goal of the study was to investigate user engagement, browsing interest and serendipity, users were always given the choice to explore the library on their own, especially when a participant found an interesting category they wanted to investigate further. 
Users were also not given a time limit for their experience and were free to explore both versions for as long as they liked.
After finishing the tasks, users were asked to fill out the post-questionnaire and repeat the procedure for the other scenario.

\section{Results}

To evaluate and compare user experience in both scenarios we considered system usability, task load, and user engagement standard questionnaire. We used T-Test for normally distributed data and Wilcoxon Signed Rank test for not-normally distributed data.

The results of the SUS questionnaire show no significant difference between VR  procedural generated library (AVG: 68.00, STD: 14.50) and traditional digital library (AVG: 66.62, STD: 17.61) with a p-value of 0.8. However, we found significant differences for individual items in SUS questionnaires. The item “I found the various functions in this system were well integrated” shows significant higher score in VR (AVG: 4.0, STD: 0.89) compared to traditional interface (AVG: 3.0, STD: 1.04) with a p-value of 0.01. with p-value of 0.02 and 0.04 correspondingly.

The users' response to task load during the experience show significantly higher "physical demand" (p-value: 0.001) in the VR environment. The other items of NASA TLX do not show significant differences. However, we can expect to see a significant difference in the "frustration" item by increasing the number of participants in the study as the current p-value is borderline (p-value: 0.055).

The UES used in this evaluation features four dimensions: Aesthetic Appeal, Focused Attention, Perceived Usability and Reward. Participants generally rated their experience to be significantly more rewarding (p-value: 1.5 $e^{-9}$) when using the VR environment. They also rated the "Aesthetic Appeal" of the VR system considerably higher (p-value: 1.1 $e^{-9}$). In addition, users of the VR system felt significantly more absorbed (p-value: 2.5 $e^{-14}$) in the experience than when using the web page. However, we do not find any significant difference in the case of "Perceived Usability" between the two systems.

As an outcome of the costume questionnaire users found the VR environment more enjoyable (p-value: 0.001) and better motivated users to browse through the presented library and explore its content (p-value: 0.0005). However, we do not find a significant difference between the two conditions in the case of ease of reading and legibility.

\section{Discussion}

Users expressed their experience in VR scenarios by overall enjoyment and worthwhile application. Users generally felt that they needed to learn more before getting going with the VR system and were more in need of assistance from a technical person. This could likely be due to the fact that users generally had little to no prior experience with VR. Despite this, they found both systems equally easy to use, and felt confident using them, given their more substantial experience with traditional computer usage.
Despite the generally positive feedback, some participants of the VR library expressed that they sometimes confused buttons and did not know which to press for a certain action.
We can expect by the increase in the spreading of consumer-based VR devices among users, they become more familiar with general interaction in this immersive environment.  In addition, having an onboarding level before asking users to enter the main VR application can help them to improve their familiarity with VR experiences.

The \textit{Physical Demand} of the VR system seems to be significantly higher, which is in line with findings from other studies, e.g. \cite{Kuber2023,safikhani2021influence}. One of the reasons for this excessive physical demand can be the difference between types of interaction in VR. In this case, mid-air interaction can lead to additional physical load. Combining different interaction types in the environment to reduce continuous mid-air interactions may help in reducing physical demands. 
Additionally, while found to be slightly over the threshold for being significant in this study, \textit{Mental Demand} is mostly likely also more demanding in VR. This can mostly be attributed to the gesture controls and the general physical effort that comes from standing and moving around as well as generally more mental strain that occurs through the usage of VR.
The most prominent differences can be seen when looking into user engagement and the results of the User Engagement Scale. Participants showed to be significantly more present in the VR experience which is generally to be expected from the usage of an immersive virtual environment \cite{slater2018,grabarczyk2016}. Furthermore, the VR environment was able to entice users’ curiosity and therefore promoted users to further investigate the displayed resources and succeeded in bringing an overall fun experience to the participants.
The significant increase in \textit{Aesthetic Appeal} in the VR version suggests that users generally thought this system was more visually pleasing. 
The VR system especially shows promising results regarding the browsing experience as participants found the system provided a smooth and seamless experience while finding the environment easy to navigate.
Users also found the interface to be significantly more motivating for browsing and felt a sense of exploration and discovery which was likely additionally contributed by the exhibition objects that participants found interesting to interact with.
Texts in both systems were rated to be similarly legible and easy to read. Book texts in the VR environment were only available as plain text and while Project Gutenberg provides well-structured text files other formats such as ePUB might lead to a better reading experience. The overall reading enjoyment was significantly higher in the VR environment.

While the study shows very promising results, future studies are needed to confirm these findings and provide more concise insights. This is especially true for longer and repeated usages of the implemented system as the current study only investigated short-term results that could be influenced by the novelty factor of the system. Additionally, the system should be evaluated against other digital library systems as well as the visuals of \textit{Project Gutenberg}'s web-interface were generally evaluated quite low. Furthermore, the browsing functionalities of the system could furthermore be expanded by providing a recommender system that suggests books based on previously read titles or categories interesting to the user. It is imperative to include a note-taking feature in future development stages to enhance reading context, especially in academic settings. Additionally, the completion model from \textit{OpenAI} could be replaced with a custom one that is trained on the actual corpus of the library and is, therefore, able to provide additional information and summaries even about newer resources. 

\section{Conclusion}
Digital libraries are innovative and dynamic information repositories that have evolved in response to the digital age. Commonly used digital library interfaces often fall short of providing a similar experience for browsing as physical libraries. Virtual reality (VR) can provide immersive and interactive experiences that already show promising results for providing increased motivation, enjoyment, and presence in other fields such as education. This study proposes an immersive VR environment that allows for the exploration and reading of books in a way that emulates the traditional library experience. The environment is generated procedurally based on a catalog of books where each room corresponds to a category of the digital library. The implemented system allows users to read through the available books in an immersive library setting. Users are furthermore provided with additional information and summaries about the books taken from a language completion model. 
To assess the benefits of the user experience using this environment, a study was conducted comparing the system against the web interface of \textit{Project Gutenberg}. The results of the study are promising as participants rated their browsing experience in VR as significantly more enjoyable and rewarding.  Additionally, the implemented system was able to convey a sense of exploration and entice the curiosity of users to continue browsing.

%
%
%
\bibliographystyle{splncs04}
\bibliography{01_Bib}

\end{document}